\documentclass[letter]{aa} 

\usepackage{graphicx}

\usepackage{txfonts}

\usepackage{booktabs}

\usepackage{natbib}

\usepackage{hyperref}  
\hypersetup{colorlinks=true,linkcolor=[rgb]{1.,0.2,0.2},citecolor=[rgb]{0.1,0.4,1.},filecolor=[rgb]{0.7,0.2,0.2},urlcolor=[rgb]{0.7,0.2,0.2}}

\newcommand{\bd}{\begin{displaymath}}
\newcommand{\ed}{\end{displaymath}}
\newcommand{\be}{\begin{equation}}
\newcommand{\ee}{\end{equation}}
\newcommand{\beaa}{\begin{eqnarray*}}
\newcommand{\eeaa}{\end{eqnarray*}}
\newcommand{\bea}{\begin{eqnarray}}
\newcommand{\eea}{\end{eqnarray}}

\def\macs1149{MACS 1149}

\begin{document} 

   \title{Augmenting the power of time-delay cosmography in lens galaxy clusters by probing their member galaxies I. Type Ia supernovae}

\author{
A.~Acebron \inst{\ref{unimi},\ref{inafmilano}} \fnmsep\thanks{E-mail: \href{mailto:ana.acebron@unimi.it}{ana.acebron@unimi.it}}
\and
S.~Schuldt \inst{\ref{unimi},\ref{inafmilano}} \and
C.~Grillo \inst{\ref{unimi},\ref{inafmilano}} \and
P.~Bergamini \inst{\ref{unimi}, \ref{inafbo}} \and
G.~Granata \inst{\ref{unimi},\ref{inafmilano}} \and
U.~Me\v{s}tri\'{c} \inst{\ref{unimi}, \ref{inafbo}}\and
G.~B.~Caminha \inst{\ref{tum}, \ref{mpa}}\and
M.~Meneghetti \inst{\ref{inafbo}} \and
A.~Mercurio \inst{\ref{salerno}, \ref{inafna}, \ref{salernoINFN}} \and
P.~Rosati \inst{\ref{unife},\ref{inafbo}} \and
S.~H.~Suyu \inst{\ref{tum}, \ref{mpa}, \ref{asiaa}}\and
E.~Vanzella \inst{\ref{inafbo}}
}
\institute{
Dipartimento di Fisica, Universit\`a degli Studi di Milano, via Celoria 16, I-20133 Milano, Italy \label{unimi}
\and
INAF -- IASF Milano, via A. Corti 12, I-20133 Milano, Italy \label{inafmilano}
\and
INAF -- OAS, Osservatorio di Astrofisica e Scienza dello Spazio di Bologna, via Gobetti 93/3, I-40129 Bologna, Italy \label{inafbo} 
\and
Technical University of Munich, TUM School of Natural Sciences, Department of Physics, James-Franck-Str.~1, 85748 Garching, Germany \label{tum}
\and
Max-Planck-Institut f\"ur Astrophysik, Karl-Schwarzschild-Str.~1, D-85748 Garching, Germany \label{mpa}
\and
Università di Salerno, Dipartimento di Fisica "E.R. Caianiello", Via Giovanni Paolo II 132, I-84084 Fisciano (SA), Italy \label{salerno}
\and
INAF -- Osservatorio Astronomico di Capodimonte, Via Moiariello 16, I-80131 Napoli, Italy \label{inafna}
\and
INFN – Gruppo Collegato di Salerno - Sezione di Napoli, Dipartimento di Fisica "E.R. Caianiello", Università di Salerno, via Giovanni Paolo II, 132 - I-84084 Fisciano (SA), Italy. \label{salernoINFN}
\and
Dipartimento di Fisica e Scienze della Terra, Universit\`a degli Studi di Ferrara, via Saragat 1, I-44122 Ferrara, Italy \label{unife}
\and
Academia Sinica Institute of Astronomy and Astrophysics (ASIAA), 11F of ASMAB, No.1, Section 4, Roosevelt Road, Taipei 10617, Taiwan \label{asiaa}
           }             


 
  \abstract{We present a simple and promising new method to measure the expansion rate and the geometry of the universe that combines observations related to the time delays between the multiple images of time-varying sources, strongly lensed by galaxy clusters, and Type Ia supernovae, exploding in galaxies belonging to the same lens clusters. By means of two different statistical techniques that adopt realistic errors on the relevant quantities, we quantify the accuracy of the inferred cosmological parameter values. We show that the estimate of the Hubble constant is robust and competitive, and depends only mildly on the chosen cosmological model. Remarkably, the two probes separately produce confidence regions on the cosmological parameter planes that are oriented in complementary ways, thus providing in combination valuable information on the values of the other cosmological parameters. We conclude by illustrating the immediate observational feasibility of the proposed joint method in a well-studied lens galaxy cluster, with a relatively small investment of telescope time for monitoring from a 2 to 3m class ground-based telescope.}

   \keywords{cosmology: observations $-$ gravitational lensing: strong $-$ supernovae: general $-$ galaxies: clusters: general $-$ methods: data analysis $-$ cosmology: cosmological parameters}
   
   \titlerunning{Combination of TDC with a SN Ia in cluster member galaxy}
   \authorrunning{Acebron et al.}
   \maketitle

%

\section{Introduction}
\label{sec:intro}

Since \citet{Refsdal1964} theoretically predicted that strongly lensed supernovae (SNe) with measured time delays between their multiple images could provide an independent way to estimate the value of the present-day cosmic expansion rate, defined as the Hubble constant ($H_0$), the time-delay cosmography (TDC) technique has been applied in several lens galaxy and galaxy-cluster systems with multiply imaged SNe or quasars (QSOs) \citep[e.g.][]{Suyu2017, Grillo2018, Birrer2019, Wong2020, Treu2022, Shajib2023}. 
The TDC method is a single-step technique (i.e. not requiring any complex calibration with distance anchors) and is completely independent from the local distance ladder and early-Universe probes. As such, this method can play a crucial role in helping to clarify the Hubble tension problem \citep[][]{Moresco2022}. 

The discovery of SN ‘Refsdal’ \citep{Kelly2015}, imaged six times by the galaxy cluster MACS J1149.5$+$2223 \citep[hereafter \macs1149,][]{Grillo2016, Treu2016, Lotz2017}, was exploited by \citet{Grillo2018, Grillo2020} to estimate the value of $H_0$ through a full strong-lensing analysis, including the measured time delays between the SN multiple images. The 6\% (statistical plus systematic) uncertainty on the value of $H_0$ achieved in \citet{Grillo2020} demonstrates that lens galaxy clusters with time delays are a valuable and complementary tool for measuring the expansion rate and the geometry of the Universe.
In this series of two Letters we examine the possibility of boosting the power of TDC in lens galaxy clusters by \textit{I)} observing Type Ia Supernovae (SNe Ia) in cluster member galaxies or \textit{II)} using the same galaxies as cosmic chronometers \citep{Bergamini2024}. 

The TDC method is a purely geometrical probe, where the cosmological dependence is fully encapsulated in the typical distances involved in a lensing system, namely in the observer-lens, the lens-source, and the observer-source angular-diameter distances. 
In this work we explore, for the first time, the idea of taking advantage of a SN Ia detected in a cluster member galaxy to provide an independent distance measurement to the lens through the distance modulus relation \citep{Riess1998, Perlmutter1999}. From the combination of these two techniques, we quantify the gain in precision on the measurement of the values of some cosmological parameters.
While about one SN Ia is expected to explode on average every century in a massive galaxy, the higher probability of observing this phenomenon in galaxy clusters, containing several hundreds of member galaxies \citep[e.g.][]{Owers2011, Rosati2014, Grillo2016, Annunziatella2017, Mercurio2021, Richard2021, Lagattuta2022, Bergamini2023a}, makes them ideal laboratories for this novel method. 

The Letter is organised as follows. In Sect.~\ref{sec:methods} we briefly describe the TDC and SNe Ia methods to measure the values of the cosmological parameters. In Sect.~\ref{sec:sims} we illustrate how we quantify the precision attainable in these measurements with the proposed combined technique. In Sect.~\ref{sec:discussion} we present our results and discuss the observational feasibility of carrying out this analysis. In Sect.~\ref{sec:conclusions} we summarise our conclusions.

In this work, magnitudes are given in the AB system.

\section{Methods}
\label{sec:methods}
In this section we concisely illustrate the dependence of some of the observables related to the multiple images of a time-variable lensed source and to SNe Ia on the values of the cosmological parameters, such as $H_0$; the present-day cosmological densities of matter, $\Omega_{\rm m}$, and of dark energy, $\Omega_{\rm de}$; and the dark energy equation-of-state parameter, $w$.

\subsection{Time-delay cosmography}
\label{sec:TDC}
Gravitational lensing occurs when the light rays from a background source are deflected by a galaxy or a galaxy cluster in the foreground, which acts as a lens. In the strong-lensing regime, multiple images of background sources are formed. If the luminosity of a multiply lensed source is intrinsically time-varying, such as that of SNe or QSOs, the differences in light arrival times between the multiple images (or time delays) can be measured \citep[e.g.][]{Fohlmeister2013, Courbin2018, Millon2020, Kelly2023}. The time delay between two images (labelled $\rm i_{1}$ and $\rm i_{2}$) of the same background source, $\Delta t_{\rm i_1 i_2}$, is
\begin{equation}
\label{eq:time_delay}
    \Delta t_{\rm i_1 i_2} = \frac{D_{\Delta t}}{c} \Delta \phi_{\rm i_1i_2} ,
\end{equation}
where $c$ is the speed of light and $\phi$ is the Fermat potential, which is related to the lens total gravitational potential. The time-delay distance, $D_{\Delta t}$ \citep{Suyu2010a}, is defined as

\be
\label{eq:time_delay_dist}
D_{\Delta t} = (1+z_\text{d}) \frac{D_\text{d}^\text{A} D_\text{s}^\text{A}}{D_\text{ds}^\text{A}},
\ee
where $z_{\rm d}$ denotes the redshift of the lens, and $D_{\rm d}^\text{A}$, $D_{\rm ds}^\text{A}$, and $D_{\rm s}^\text{A}$ are the angular-diameter distances between the observer and the lens, the lens and the source, and the observer and the source, respectively.
The cosmological dependence is embedded in the time-delay distance through the ratio of these three angular-diameter distances. This term can thus be expressed as a function of the redshifts of the lens and the source, $z_{\rm s}$, and depends on the values of the cosmological parameters $D_{\Delta t}(z_{\rm d}, z_{\rm s}; H_0, \Omega_{\rm m}, \Omega_{\rm de}, w)$.
As shown by the Sobol’ sensitivity analysis \citep{Sobol2001}, the time-delay distance is primarily sensitive to the value of $H_0$ (as $D_{\Delta t} \propto H_0^{-1}$), and more mildly to those of $\Omega_{\rm m}$, $\Omega_\text{de}$, and $w$ \citep[see Fig. 24 in][]{Moresco2022}.

As can be seen from Eq.~(\ref{eq:time_delay}), the uncertainty on the value of $D_{\Delta t}$ (and therefore on that of $H_0$) is approximately the sum in quadrature of the uncertainties on the time-delay measurement and on the lens total mass distribution. 
Time delays in lens clusters can be long (i.e. more than a year) and can thus be measured with a relative precision better than $\sim2\%$ \citep{Fohlmeister2013, Dahle2013, Munoz2022}. In this case the error budget on $D_{\Delta t}$ is dominated by the uncertainty associated with the total mass distribution of the lens galaxy cluster.
The relative error on the value of $D_{\Delta t}$ achieved from a single (galaxy or cluster) strong-lensing system ranges typically from $\sim4\%$ to $\sim9\%$ \citep[see][]{Suyu2014, Wong2017, Bonvin2017, Grillo2018, Birrer2019, Chen2019, Grillo2020, Rusu2020, Shajib2020, Wong2020, Shajib2023}.

\begin{figure*}[ht!]
    \centering
    \includegraphics[width=\columnwidth]{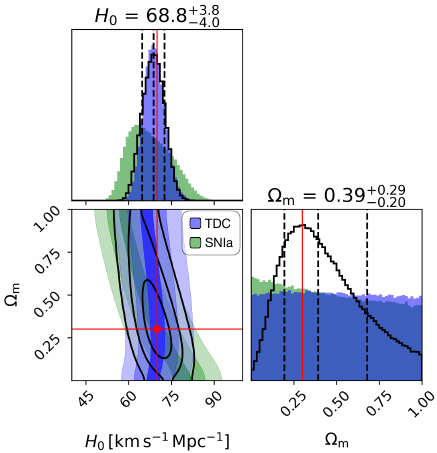}
    \includegraphics[width=\columnwidth]{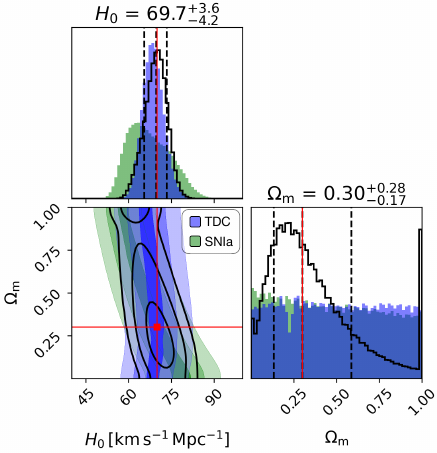} 
    \caption{Inference from the Bayesian (\textit{left}) and Monte Carlo (\textit{right}) approaches on the values of the cosmological parameters $H_0$ and $\Omega_\text{m}$ in a flat-$\Lambda$CDM model, assuming a total relative uncertainty of 5\% for both cosmological probes (with fixed $z_{\rm d}=0.54$, $z_{\rm s}=1.49$, and $z_{\rm SNIa}=0.54$). The marginalised posterior distributions for the TDC and the SN Ia luminosity distance methods are shown in blue and green, respectively. The combined distributions are shown in black. The 16$\text{th}$, 50$\text{th}$, and 84$\text{th}$ percentiles of the combined marginalised distributions are highlighted with vertical dashed lines, and the corresponding values are reported. The fiducial values are in red. The contour levels on the planes represent the 1, 2, and 3$\sigma$ confidence regions.}
    \label{fig:Bayesian-2params}
\end{figure*}

\subsection{Type Ia supernovae}
\label{sec:SNIa}

A SN Ia occurs when a carbon-oxygen white dwarf accretes enough matter from a companion star to approach the Chandrasekhar mass limit or through the merging of two white dwarfs \citep[see e.g.][for a review]{Liu2023}.
SNe Ia have been found to be a highly homogeneous population, with peak absolute magnitudes in their light curves showing a small intrinsic scatter \citep[e.g.][]{Branch1998, Freedman2010}.
These sources can be exploited as standardisable candles, after applying an empirical calibration that depends on the decline rate in the 15 days following the maximum light \citep{Pskovskii1984, Phillips1993}.
Being very bright objects, SNe Ia are powerful distance indicators out to high redshifts \citep{Riess2018}. The distance to a SN Ia is determined through the relation
\be
\label{eq:distmodulus}
\mu=m-M=5 \log_{10}\left(\frac{D^\text{L}}{{\rm 1~Mpc}} \right)+25, 
\ee
where $\mu$ is known as the distance modulus, and $m$ and $M$ denote the apparent (corrected for interstellar dust extinction) and absolute magnitudes of the SN Ia, respectively. The luminosity distance of a SN Ia, $D^\text{L}$, can be expressed as a function of its redshift, $z_{\rm SNIa}$, and of the values of the cosmological parameters, $D^\text{L}(z_{\rm SNIa}; H_0, \Omega_{\rm m}, \Omega_{\rm de}, w)$.
As is true for $D_{\Delta t}$, $D^\text{L}$ is proportional to the value of $H_0^{-1}$ and is less sensitive to the values of the other cosmological parameters. By computing the value of the Sobol’ index at different redshifts, we find that $\sim100\%$ to $\sim60\%$ of the variance of the luminosity distance is due to the variance of $H_0$ from $z_{\rm SNIa}=0$ to $z_{\rm SNIa}=1$, respectively.

The total error budget on the value of the luminosity distance of a single SN Ia includes the contributions from several factors, such as photometric errors, light-curve fitting, galaxy host and dust extinction corrections, and colour variation \citep[see e.g.][]{Betoule2014, Scolnic2014, Scolnic2022}. With this method the individual distances to SN Ia host galaxies can be measured to a precision of 5\% to 10\% \citep{Riess2022}.
These uncertainties include the error contribution from the geometric anchors used by the SH0ES team \citep[see][]{Riess2022}.
Alternatively, the quantity $D_{\Delta t}$, measured through the TDC technique for each considered lens galaxy cluster, can be used as a geometric calibrator through an inverse distance ladder approach \citep[see][]{Taubenberger2019}. The analysis would then be fully self-consistent and independent of the local distance ladder, although possibly with larger uncertainties on the measurement of $D^\text{L}$ depending on the number of anchors and their individual uncertainties.

\section{Simulations}
\label{sec:sims}

In this section we describe two different approaches, a Bayesian and a Monte Carlo method, used to explore the precision with which the combined technique described above can probe the cosmological parameters. 
We checked that the results obtained with the two approaches are consistent, and, for brevity, we focus mainly on the first approach.

\subsection{Assumptions}
\label{sec:assumptions}

As demonstrated in \citet{Grillo2018, Grillo2020}, \macs1149 offers excellent opportunities for cosmological applications. Thus, we chose to consider as a reference this lens galaxy cluster, located at $z_{\rm d}=0.54$ \citep{Grillo2016}. \macs1149 hosts SN Refsdal, a strongly lensed SN at $z_{\rm s}=1.49$ with measured time delays between five of its multiple images \citep{Kelly2015, Kelly2016, Kelly2023}, together with 84 additional multiple images from 27 other background sources \citep{Grillo2016}.
If a SN Ia were detected in one of the cluster members of \macs1149, one would have that $z_{\rm SNIa}=z_{\rm d}=0.54$ \citep{Leget2018}.
 
As quoted in Sects.~\ref{sec:TDC} and \ref{sec:SNIa}, the typical total (i.e. statistical plus systematic) relative errors currently obtained for the time-delay distance, $\sigma_{D_{\Delta t}}$, and the SNe Ia luminosity distance, $\sigma_{D_{\text{d}}^{\rm L}}$, both range from approximately 5\% to 10\%. In this work we adopted these two values to mimic more conservative and less conservative scenarios. 

Finally, we performed our analysis in a flat-$\Lambda$CDM ($\Omega_{\rm m} + \Omega_{\rm de}= 1$ and $w=-1$) and an open-$w$CDM cosmological model, described by two and four free parameters, respectively. We assumed a fiducial cosmology with $H_0 = 70~\mathrm{km~s^{-1}~Mpc^{-1}}$, $\Omega_{\rm m}~=~0.3$, $\Omega_{\rm de}~=~0.7$, and $w=-1$, and we considered large uniform priors ($H_0 \in [20, 120]~\mathrm{km~s^{-1}~Mpc^{-1}}$, $\Omega_{\rm m}$ and $\Omega_{\rm de} \in [0, 1]$, and $w\in [-2, 0]$) on the free cosmological parameters.

\subsection{Bayesian approach}
\label{sec:sims:bayesian}
The values of the cosmological parameters and their uncertainties are estimated by sampling a log-posterior, as described below. 
The likelihood function, $\mathcal{L}_i$, for a given technique $i$ is defined as  
\be
\label{eq:lnLK_eq}
\mathcal{L}_i = \frac{1}{\sigma_i \sqrt{2\pi}}\exp(-\chi^2_i/2),
\ee
assuming that the errors associated with the measurements, $\sigma_i$, are Gaussian. The term $\chi^2_i$ denotes the chi-square function of each technique, which can be expressed as

\be
\label{eq:chi2_deltat}
\chi^2_{\rm TDC} = \left(\frac{D_{\Delta t, ~\rm true} - \hat{D}_{\Delta t}}{\sigma_{D_{\rm \Delta t, ~\rm true}}}\right)^2 
\ee
for the TDC method, and as

\be
\label{eq:chi2_SNIa}
\chi^2_{\rm SNIa} = \left(\frac{D^{\rm L}_{\rm true} - \hat{D}^\text{L}}{\sigma_{D^{\rm L}_{\rm true}}}\right)^2 
\ee
for the SN Ia luminosity distance method. The quantities $D_{\Delta t, ~\rm true}$ and $D^{\rm L}_{\rm true}$ correspond to the values of the time-delay and the luminosity distances, respectively, computed in the fiducial input cosmological model, and $\hat{D}_{\Delta t}$ and $\hat{D}^\text{L}$ are the values of these distances obtained by sampling the two-dimensional ($H_0, ~\Omega_{\rm m}$) or four-dimensional ($H_0, ~\Omega_{\rm m}, ~ \Omega_{\rm de} , ~ w$) parameter space (with fixed $z_{\rm d}=0.54$, $z_{\rm s}=1.49$, and $z_{\rm SNIa}=0.54$).
The total posterior probability distribution is then obtained by multiplying the values of the likelihood from each technique ($\mathcal{L}_\text{tot} = \mathcal{L}_{\rm TDC} \times \mathcal{L}_{\rm SNIa}$) and assuming the priors detailed in Sect.~\ref{sec:assumptions} for the cosmological parameters.

To sample the posterior distribution of the cosmological parameter values, we exploited the Affine-Invariant Markov chain Monte Carlo (MCMC) Ensemble sampler developed by \citet{Goodman-Weare2010}, and in particular its Python implementation \citep{emcee}. The parameter space is explored with ten walkers, with $10^5$ steps each.
We removed the first 5000 steps of each walker as the burn-in phase, ensuring that the MCMC chains have converged and that the results are independent of the initial position of the walkers. This number of steps is $\sim$100 times larger than the integrated auto-correlation time computed for each parameter. 

\begin{table*}[ht!]
  \renewcommand\arraystretch{1.4}
  \centering
  \caption{Intervals at the 68\% confidence level for the values of the cosmological parameters obtained with the Bayesian method.}
  \label{table:1}
  \begin{tabular}{cccccccc}
    \hline
    \hline
    & &
    \multicolumn{2}{c}{flat-$\Lambda$CDM$^{\mathrm{a}}$} &
    \multicolumn{4}{c}{open-$w$CDM} \\
    \cmidrule(r){3-4}\cmidrule(l){5-8}
Err.$^\mathrm{b}$ $D_{\Delta t}$ & Err.$^\mathrm{b}$ $D^\text{L}$ & $H_0$$^{\mathrm{c}}$ & $\Omega_{\rm m}$ & $H_0$$^{\mathrm{c}}$ & $\Omega_{\rm m}$ & $\Omega_{\rm de}$ & $w$ \\
    \hline
$5\%$ & $5\%$ & $68.8^{+3.8}_{-4.0}$ & $0.39^{+0.29}_{-0.20}$ & $68.5^{+6.4}_{-5.1}$ & $0.47^{+0.35}_{-0.32}$ & $0.63^{+0.24}_{-0.32}$ & $-1.12^{+0.63}_{-0.60}$ \\
$5\%$ & $10\%$ & $68.8^{+3.8}_{-3.7}$ & $0.45^{+0.33}_{-0.26}$ & $69.1^{+6.0}_{-5.1}$ & $0.47^{+0.35}_{-0.32}$ & $0.55^{+0.30}_{-0.35}$ & $-1.06^{+0.67}_{-0.64}$ \\
$10\%$ & $5\%$ & $67.9^{+6.2}_{-5.7}$ & $0.42^{+0.34}_{-0.25}$ & $67.0^{+7.7}_{-5.6}$ & $0.48^{+0.35}_{-0.33}$ & $0.57^{+0.29}_{-0.35}$ & $-1.09^{+0.67}_{-0.63}$ \\
$10\%$ & $10\%$ & $68.7^{+6.4}_{-5.8}$ & $0.45^{+0.34}_{-0.28}$ & $68.5^{+7.9}_{-6.3}$ & $0.47^{+0.35}_{-0.33}$ & $0.54^{+0.31}_{-0.35}$ & $-1.07^{+0.68}_{-0.65}$ \\ 
 \hline
    \hline
  \end{tabular}
  \begin{list}{}{}
\item[] 
$^{\mathrm{a}}$ $\Omega_{\rm m}+\Omega_{\rm de}=1$ and $w=-1$. \\
$^{\mathrm{b}}$ Adopted percentage relative errors. \\
$^{\mathrm{c}}$ ($\mathrm{km~s^{-1}~Mpc^{-1}}$).
\end{list}
\end{table*}

\subsection{Monte Carlo approach}
\label{sec:sims:grid}

In this case we first computed the values of the time-delay, $D_{\Delta t, \text{true}}$, and luminosity, $D^{\rm L}_\text{true}$, distances assuming the fiducial cosmology (and $z_{\rm d}=0.54$, $z_{\rm s}=1.49$, and $z_{\rm SNIa}=0.54$). 
Then, from Gaussian distributions centered on these values and with standard deviations equal to the uncertainties reported in Sect.~\ref{sec:assumptions}, we extracted $10^6$ time-delay and luminosity distance values, which represent our possible measurements. 
Next, we built a two-dimensional ($H_0, ~\Omega_{\rm m}$) and a four-dimensional ($H_0, ~\Omega_{\rm m}, ~ \Omega_{\rm de} , ~ w$) grid for the flat-$\Lambda$CDM and open-$w$CDM models, respectively, covering with 1000 bins of equal width the assumed prior intervals (see Sect.~\ref{sec:assumptions}) of the cosmological parameters. For every grid point, we computed the values of $D_{\Delta t}$ and $D^\text{L}$ with the corresponding cosmological parameters.
Finally, for each of the $10^6$ simulated measurements, we
searched for the best-fit values of the cosmological parameters by minimising a total chi-square function, $\chi^2_\text{tot}$, defined as 
\be
\label{eq:chi2_tot}
\chi^2_\text{tot} = \chi^2_{\rm TDC} + \chi^2_{\rm SNIa},
\ee
where the expressions for $\chi^2_{\rm TDC}$ and $\chi^2_{\rm SNIa}$ are given, respectively, in Eqs.~(\ref{eq:chi2_deltat}) and (\ref{eq:chi2_SNIa}), and the considered errors for the distances correspond to 5\% or 10\% of the sampled quantities.

\section{Discussion}
\label{sec:discussion}
In Table~\ref{table:1} we summarise the median values and the 1$\sigma$ confidence level intervals for the values of the cosmological parameters within the chosen cosmological models, assuming different values for the relative uncertainties on $D_{\Delta t}$ and $D^\text{L}$. 
Figure~\ref{fig:Bayesian-2params} shows in a flat-$\Lambda$CDM model the posterior probability distributions and the 1, 2, and 3$\sigma$ confidence regions for $H_0$ and $\Omega_{\rm m}$ inferred from the TDC (blue), the SN Ia luminosity distance (green), and their combination (black), when assuming a 5\% relative error for both cosmological probes.
We observe that the intrinsic degeneracy between $H_0$ and $\Omega_{\rm m}$ from the TDC and the SN Ia luminosity distance methods are oriented in slightly different directions, making these probes complementary \citep[e.g.][]{Moresco2022}. 
As expected (see Sect.~\ref{sec:methods}), both techniques are more sensitive to the value of $H_0$ than to that of $\Omega_{\rm m}$. The value of $\Omega_{\rm m}$ cannot be measured from either the TDC or the SN Ia luminosity distance method alone, as found for example in single strong-lensing systems \citep[see][]{Suyu2010, Bonvin2017, Wong2017, Birrer2019}. Interestingly, the combination of the two techniques results in an estimate of the value of $\Omega_{\rm m}$, however with a quite significant statistical error. 
When considering lens galaxy clusters, we note that the observed positions of a large number of multiple images at different redshifts provide information about the `family ratios', from which the values of $\Omega_{\rm m}$, $\Omega_{\rm de}$, and $w$ can also be inferred \citep[as shown in][]{Soucail2004, Jullo2010, Linder2011, Caminha2016, Acebron2017, Grillo2018, Caminha2022}. For simplicity, in this pilot study we neglected the contribution of the family ratio term, thus obtaining conservative estimates of the cosmological parameter values.
We also note that the uncertainty on $H_0$ is driven by that of the $D_{\Delta t}$ term (see Table~\ref{table:1}). For a fixed value of $\sigma_{D_{\Delta t}}$ a measurement of the SN Ia luminosity distance with a 5\% or 10\% error results in a similar precision on the measurement of $H_0$. Nevertheless, a measurement of $D^\text{L}$ nicely complements and enhances the TDC technique. In particular, for lens clusters with a 10\% relative uncertainty both on $D_{\Delta t}$ and $D^\text{L}$, the joint method enables a gain in precision on the $H_0$ estimate by a factor of $\sim 1.2$ compared to the results from the TDC technique alone.

\begin{figure}
    \centering
\includegraphics[scale=2]{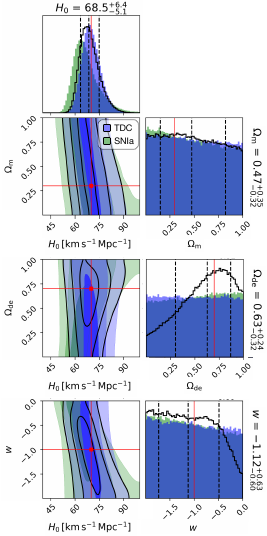}
    \caption{Posterior probability distributions from the Bayesian approach of $H_0$, $\Omega_{\rm m}$, $\Omega_{\rm de}$, and $w$ in an open-$w$CDM model, assuming a total, relative uncertainty of 5\% for both cosmological probes (with fixed $z_{\rm d}=0.54$, $z_{\rm s}=1.49$, and $z_{\rm SNIa}=0.54$). The colour-coding and shown statistical quantities are as in Fig.~\ref{fig:Bayesian-2params}.}
    \label{fig:Bayesian-4params}
\end{figure}

In Fig.~\ref{fig:Bayesian-4params} we show the posterior probability distributions and the 1, 2, and 3$\sigma$ confidence regions for $H_0$, $\Omega_{\rm m}$, $\Omega_{\rm de}$, and $w$ in an open-$w$CDM cosmology, assuming a 5\% relative error for both probes. This figure illustrates that the value of $H_0$ is robustly measured, almost independently of the assumed cosmological models, with a posterior probability distribution slightly larger here than in a flat-$\Lambda$CDM model. As in that model, the probability distribution functions of the other cosmological parameters are quite flat from the TDC or SN Ia luminosity distance method alone, but their combination leads to more precise measurements (see also Table~\ref{table:1}). For instance, values of $\Omega_{\rm m} > \text{0.89}$, $\Omega_{\rm de} < \text{0.21}$, and $w > -\text{0.36}$ are ruled out at the 90\% confidence level.

After illustrating the possibility of boosting the TDC method in combination with a SN Ia luminosity distance measurement in a cluster member galaxy, we discuss the observational feasibility of carrying out this new method.
As previously mentioned, the probability of observing a SN Ia in a cluster galaxy, as estimated from several observational programmes, is very low \citep[i.e. $\gtrsim 0.1 \times 10^{-12}$ SNe $\rm M_{\odot}^{-1}$ yr$^{-1}$, see e.g.][]{Sharon2010, Dilday2010, Petrushevska2016, Toy2023, Golubchik2023}. However, lens galaxy clusters host hundreds to thousands of member galaxies, often possessing extensive high-quality spectroscopic observations \citep[e.g.][]{Braglia2009, Owers2011, Rosati2014, Grillo2016, Mercurio2021, Lagattuta2022}. For instance, in \macs1149 a total of 195 spectroscopically confirmed cluster galaxies have been detected over an area of $\sim$ 6.7 arcmin$^2$ (Schuldt et al. in prep.). The Andalucia Faint Object Spectrograph and Camera (ALFOSC) at the 2.5m Nordic Optical Telescope (NOT) on La Palma (Spain) was already used to search for high-redshift SNe in lens cluster fields \citep[see e.g.][]{Petrushevska2016}. The ALFOSC instrument has a field of view of 6.4 arcmin across. This side corresponds to approximately 2.4 Mpc at the redshift of \macs1149. Several studies of lens galaxy clusters, with virial masses similar to that of the cluster used as a reference here, have shown that the total stellar mass enclosed within a circle with radius equal to 300 kpc ranges from 2 to 3 $\times$10$^{12}$ M$_{\odot}$ (see Fig.~3 in \citealt{Annunziatella2017} and Fig.~9 in \citealt{Granata2022}). By extrapolating these values to a radius of 1.2 Mpc or considering the projected stellar mass density profile shown in Fig.~12 by \citet{Annunziatella2014} and integrating it within the same radius, we estimated a cumulative stellar mass value of $\gtrsim 10^{13}$ M$_{\odot}$ in the member galaxies of a massive galaxy cluster like \macs1149. This translates into a count rate of $\gtrsim~1$ SN Ia yr$^{-1}$. This estimate is conservative as the rate of SNe Ia exploding in a lens cluster would be higher in a field of view larger than the relatively small ALFOSC field, or by targeting higher-redshift galaxy clusters, as shown by the results from the HST Cluster SN Survey \citep[P.I. Perlmutter,][]{Dawson2009, Barbary2012}.
Finally, the typical SN~Ia peak magnitude of $M(B)=-19.5$ mag in the $B$ band corresponds to $ m(B)\sim 23$ mag (rest-frame) for a SN Ia at the redshift of \macs1149.
The detection of a SN~Ia several days before its maximum brightness (e.g. when $m(B)\sim~25$~mag) is within the capabilities of a 2 to 3m class ground-based telescope, such as the NOT, within about one hour of integration time \citep{Petrushevska2016}. This demonstrates the possibility of successfully detecting and monitoring one such event in a limited amount of time of a dedicated programme and, ultimately, of applying the proposed joint method.

\section{Conclusions}
\label{sec:conclusions}

In this work we have presented, for the first time, how to enhance the power of time-delay cosmography in lens galaxy clusters by detecting Type Ia supernovae in member galaxies, allowing for an additional and independent measurement of the luminosity distance to its host galaxy, and thus to the lens cluster. 

Using as a reference the lens galaxy cluster MACS J1149.5$+$2223, hosting the strongly lensed SN ‘Refsdal’, we have examined the complementarity of the two techniques and quantified, through their combination, the precision attainable in the measurements of the most relevant cosmological parameters.
The novel combined method provides promising predictions and can in principle be applied to any lens galaxy cluster with multiple images of a time-varying source \citep[e.g.][]{Inada2003, Oguri2010, Dahle2013, Acebron2022b, Acebron2022a, Martinez2023, Napier2023b}.
Forthcoming  , performed by the Vera C. Rubin Observatory (Legacy Survey of Space and Time) and the Euclid satellite, are expected both to significantly increase the sample of these cluster-scale strong-lensing systems and likely to detect more than one SN Ia exploding in the member galaxies of the same cluster. In the latter case the proposed technique would become even more powerful. We have shown that all the observations needed to obtain the first (and future) results with this joint method are feasible with a modest investment of ground-based telescope time.

In the second Letter of this series \citep{Bergamini2024}, we explore the possibility of analysing extensive high-quality spectro-photometric datasets (already available) in several lens galaxy clusters to homogeneously select pure samples of red, massive, and passive cluster members and exploit them as cosmic chronometers. 
By measuring the age of these objects in different lens clusters located in close-by redshift bins, it is possible to probe the expansion history of the Universe, $H(z)$, at the effective redshift of the considered lens clusters \citep[see e.g.][]{Jimenez2002, Stern2010a, Moresco2012a}, complementing the results obtained with the time-delay cosmography tecnique.

\begin{acknowledgements}
We kindly thank the anonymous referee for the useful suggestions received.
AA has received funding from the European Union’s Horizon 2020 research and innovation programme under the Marie Skłodowska-Curie grant agreement No 101024195 — ROSEAU. 
We acknowledge financial support through grants PRIN-MIUR 2017WSCC32 and 2020SKSTHZ. PR acknowledges FIR 2021 fund. SHS thanks the Max Planck Society for support through the Max Planck Fellowship.
This research was supported by the Munich Institute for Astro-, Particle and BioPhysics (MIAPbP) which is funded by the Deutsche Forschungsgemeinschaft (DFG, German Research Foundation) under Germany's Excellence Strategy – EXC-2094 – 390783311. 
Software Citations:
This work uses the following software packages:
\href{https://github.com/astropy/astropy}{\texttt{Astropy}}
\citep{astropy1, astropy2},
\href{https://github.com/dfm/corner.py}{\texttt{Corner.py}}
\citep{corner},
\href{https://github.com/dfm/emcee}{\texttt{Emcee}}
\citep{emcee},
\href{https://github.com/matplotlib/matplotlib}{\texttt{matplotlib}}
\citep{matplotlib},
\href{https://github.com/numpy/numpy}{\texttt{NumPy}}
\citep{numpy1, numpy2},
\href{https://www.python.org/}{\texttt{Python}}
\citep{python},
\href{https://github.com/scipy/scipy}{\texttt{Scipy}}
\citep{scipy}.
\end{acknowledgements}

%
%

\bibliographystyle{aa} 
\bibliography{refs} 

\end{document}